\documentclass[aps,twocolumn,prl,showpacs]{revtex4}
\usepackage{graphicx}
\usepackage{amsfonts}
\usepackage{amssymb}
\usepackage{amsmath}

\begin{document}

\title{Equation-free dynamic renormalization in a glassy compaction model}

\author{L. Chen$^1$, I.G. Kevrekidis$^{1,2}$ and P.G. Kevrekidis$^3$}
\affiliation{
$^1$Department of Chemical Engineering and $^2$PACM, 
Princeton University, 
6 Olden Str.\ Princeton, NJ 08544 USA, \\
$^3$Department of Mathematics and
Statistics, University of Massachusetts,
Amherst MA 01003-4515, USA}

\begin{abstract}
Combining dynamic renormalization with equation-free computational
tools, we study the apparently asymptotically self-similar evolution
of void distribution dynamics in the diffusion-deposition problem
proposed by Stinchcombe and Depken [Phys. Rev. Lett. {\bf 88}, 125701
(2002)].  We illustrate fixed point and dynamic approaches, forward as
well as backward in time.
\end{abstract}

\maketitle

The study of glasses is an important topic under
intense investigation for several decades now (see e.g., the review
of \cite{pablo}). 
Experimental studies in granular compaction (such as the
ones in \cite{chicago}) have offered considerable
insights on the behavior and temporal dynamics of glass forming systems. 
From the theoretical point of view, a variety of
modeling/computational approaches have been used to understand
better the key observables and their evolution for
this slow relaxational type of dynamics. 
While direct (molecular dynamics) simulations are widely used, 
perhaps more popular are
kinetic approaches including e.g., the Fredrickson-Anderson model
\cite{fa} and more recent variants thereof \cite{sollich}.
Another common approach involves the use of mode coupling theory
\cite{g67} examining the time evolution of the decay of density
fluctuations to separate liquid from glassy (non-ergodic due to
structural arrest) dynamics. 
More recently, the energy landscape of glass forming systems and the 
role of its ``topography'' has become the focus of numerous works
\cite{sciortino}.

In a previous  paper \cite{PLA} we proposed a simple compaction 
``thought experiment'' for hard spheres: the insertion of
a hard sphere in a gas of hard spheres (accepted when the sphere
does not overlap with previously existing ones).
Combining  simple thermodynamic arguments with equilibrium distribution results
we argued that the evolution of the hard sphere density 
should be logarithmic in time (as the maximal density is approached).
%
%{\bf Based on the same arguments, combined with the 
%considerations of \cite{torquato}, the conditional pair distribution 
%function $G$ (a relevant ``statistical measure'' of the system) was found 
%to decay with distance self-similarly as $1/r$ and
%its amplitude to grow in time logarithmically.
%PK: I thought that this could be easily cut but feel free to bring it back.}
% I also commented out Torquato's reference with it.

In a recent paper \cite{Stinchcombe} Stinchcombe and Depken presented 
an interesting 
diffusion-deposition model exhibiting ``glassy'' compaction kinetics
(see also \cite{physd} for a similar model example).
The model provides a useful paradigm for testing the hypothesis
of self-similar evolution of the system statistics --in
particular, of the density and the void distribution function--
based on direct simulations.

The techniques that we will use to test this hypothesis 
combine template-based dynamic renormalization \cite{template} with the
so-called ``equation-free'' approach to complex/multi-scale system modeling
\cite{Manifesto,JNNFM}.
Dynamic renormalization has been developed in the context of
partial differential equations with self-similar solutions
(e.g. blowups in finite time) \cite{Papanicolaou}.
%{\bf Panagioti, choose a couple}.
%
The equation-free approach targets systems for which a fine scale,
atomistic/stochastic simulator is available, but for which no
closed form coarse-grained, macroscopic evolution equation has
been derived.
The main idea is to substitute function and derivative evaluations
of the unavailable coarse-grained evolution equation with
short bursts of appropriately initialized fine scale computation.
The quantities required for traditional continuum numerical
analysis are thus estimated {\it on demand} from brief computational 
experimentation with the fine scale solver.
We will demonstrate equation-free fixed point approaches to computing
the self-similar shape and dynamics of the void cumulative
distribution function.

We start with a brief description of the  model and 
key (for our purposes) results of \cite{Stinchcombe}. 
We then perform
equation-free dynamic renormalization computations for this 
problem. 
%using both time evolution and fixed point methods.
%
The coarse-grained description of the system evolution backward in
time is examined through reverse coarse projective integration.
We conclude with a brief discussion of potential extensions
of the approach.

The model considered by Stinchcombe and Depken
\cite{Stinchcombe} consists of unit-size grains interacting 
through hard-core potentials and performing a Monte Carlo random walk.
While some of their results apply in any dimension,
in this paper we will work in one spatial dimension;
this is for convenience--our approach is not limited to 1d.
While the grains diffuse on the line, as soon as a 
sufficiently large void is formed, it is instantaneously
filled by an additional grain.
We will work with a system of finite size and 
periodic boundary conditions; the system size L (here of
the order of $10^5$) is
large compared to the grain size, which is taken to be 1.
If  $\rho$ is the system density,
$\delta = (1-\rho)/\rho$, is the average void size.
One of the main findings in \cite{Stinchcombe} is that
the void density $\epsilon = 1 - \rho$ 
goes to zero as an inverse logarithm 
\begin{equation}
\langle \epsilon(t) \rangle \sim \frac{1}{\ln(t)},
\label{eveq0}
\end{equation}
asymptotically for large $t$. 
A similar asymptotic behavior follows for $\delta$.

\begin{figure}[htbp]
\centering
\leavevmode
\includegraphics[height = 2.5in]{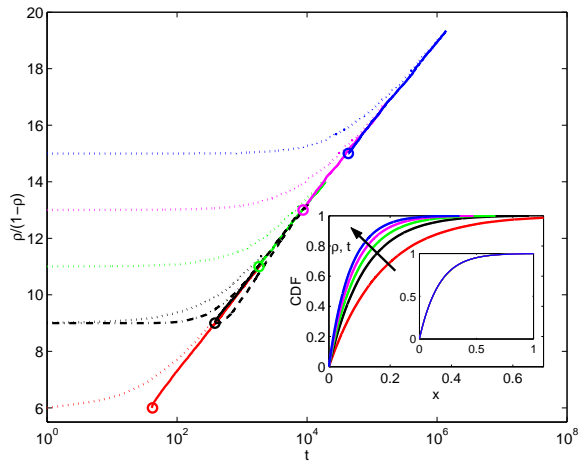}
\caption{Dynamic evolution of the model for different initializations.
Dotted (resp. dash-dotted) lines: starting with the self-similar
(resp. uniform) void CDF.
Solid (resp. dashed) lines: evolution replotted in terms of an
appropriate ``initial time" $t_0$; notice the collapse of the curves
on a straight line, see text.
The inset shows void CDFs at different time/density instances
(color-coded as in the main Figure); they all collapse upon rescaling
to a single CDF (smaller inset).
}
\label{fig1}
\end{figure}

\begin{figure}[htbp]
\centering
\leavevmode
\includegraphics[height = 1in]{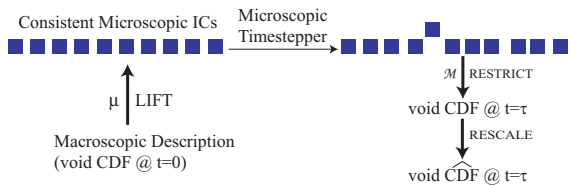}
\caption{Schematic of coarse dynamic renormalization.}
\label{fig2}
\end{figure}

Density provides a cumulative measure of the system evolution
(it is the zeroth moment of the particle distribution function).
We set out to study in detail the evolution of the
void cumulative distribution function, $CDF(x,t)$.
This function is defined between $0$ and $1$ in the void size $x$, since at 
no time do we get
voids larger than $1$ (the moment they appear, they get filled).
The dotted lines (and the dash-dot line) in Fig. \ref{fig1} show
the evolution of $\delta(t)$ for various initial void distributions (the
choice of initial void distributions for these transients will be
discussed below).
The stochastic simulation uses 100,000 particles;
sequential update is used, with a maximum spatial step size of $0.5$.
The semilog plot clearly indicates the asymptotic logarithmic
regime in time. 
The inset shows the shape of the void $CDF$ for various
simulations and various instances in time, 
%more or less
within the logarithmic regime; all shapes
when appropriately rescaled with the average void size
collapse on a single curve (smaller inset).
These two observations (logarithmic time dependence, and
collapse of the distributions upon rescaling) clearly 
suggest the possibility of self-similar evolution dynamics.

In particular, the graph of the time evolution of the densities 
suggests that $\delta$ becomes eventually proportional 
%up to an amplitude rescaling, 
to
\begin{equation}
\delta \equiv \frac{1-\rho}{\rho} \sim \frac{1}{A {\rm ln}(t+t_0) + B}.
\label{delta}
\end{equation}
Replotting the data in view of the above relation collapses the
Fig. \ref{fig1} curves onto a single line; the role of $t_0$ will be
discussed below.
This implies that the dynamical evolution of $\delta$ eventually follows
the differential equation:
\begin{equation}
\frac{d \delta}{dt} \sim -\delta^2 \exp \left(-\frac{1}{A \delta}\right).
\label{eveq1}
\end{equation}
(and hence a similar equation can easily be obtained for 
$\epsilon$). 
%while the asymptotic considerations of
%\cite{Stinchcombe} yield 
%\begin{eqnarray}
%\frac{d \langle \epsilon \rangle}{dt} \sim 
%\frac{1}{\langle \epsilon \rangle^3} \exp \left(-
%\frac{1}{\langle \epsilon \rangle}\right).
%\label{eveq2}
%\end{eqnarray}
%
The dynamics of (\ref{eveq1}) share the asymptotic
behavior of Eq. (\ref{eveq0}).
%
%The actual exact equation is:
%\begin{eqnarray}
%\frac{d \delta}{dt} = -A \exp\left(\frac{B}{A}\right) 
%\delta^2 \exp \left(-\frac{1}{A \delta}\right)
%\label{eveq1a}
%\end{eqnarray}
%(which of course reduces to the equations we have used when $A=1$ and
%$B=0$).

\begin{figure}[htbp]
\centering
\leavevmode
\includegraphics[height = 2in]{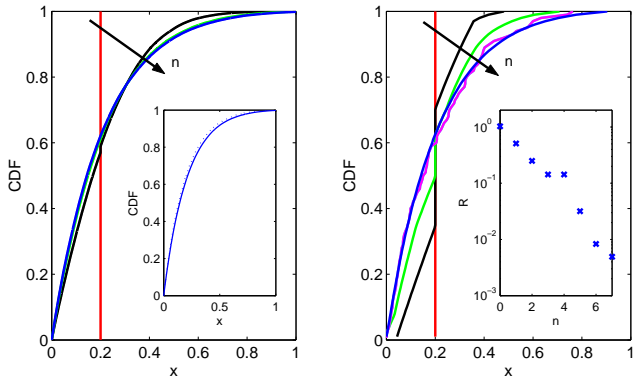}
\caption{a. Successive evolution of an initially uniform CDF profile with
renormalization: the inset shows the profile
at the n=3 iteration before (dashed) and after (solid) rescaling. 
b. Newton-Krylov-GMRES fixed point renormalization calculation, starting from
uniform void CDF; the inset shows the decrease of the norm of the residual vector
at each Newton step.}
\label{fig3}
\end{figure}

We now test the self-similarity hypothesis for the $CDF$ evolution.
If the dynamics of the $CDF$ are self-similar and stable,
all initial distributions will asymptotically approach
the same self-similar shape (modulo rescaling)
while they densify and slow down.
Our goal is to find the self-similar shape at a {\it convenient}
scale -- a relatively low density, when the evolution dynamics
are still fast.
Our main assumption is that the void $CDF$ is a good observable 
for the system dynamics -- i.e., that a macroscopic evolution equation 
(possibly averaged over several experiments) conceptually exists for
its dynamics, even though it is not available in closed form.
Starting with a given void $CDF_0$
with $\delta = \delta_0$ we {\it lift} to an ensemble
of microscopic realizations of it -- that is, 
grain configurations possessing the given void $CDF$.
This ``lifting'' is accomplished here
by randomly selecting the voids from $CDF_0$ as grains
are ``put down'' on the line; $\delta_0$ indirectly selects the scale 
at which we will perform our computation.
We then evolve each of these realizations based on {\it true}
system stochastic dynamics for a time horizon $\tau$.
Finally, we obtain
the ensemble averaged void $CDF(x,\tau)$ and its $\delta(\tau)=\delta_1$.
This is the {\it restriction} step of equation-free computation: 
evaluating the macroscopic observables of 
detailed, fine scale computations.
We now {\it rescale} our macroscopic observable (the $CDF(x,\tau)$), using
%as a template 
the ensemble average $\delta_1$, to the original $\delta_0$:
$\widehat{CDF}(\delta_0 x/ \delta_1, \tau) =  CDF(x, \tau)$.
(Clearly this requires the largest void in the
rescaled $\widehat{CDF}$ to be less than $1$ -- a condition that
one may expect to prevail at high enough densities).
We then {\it discard} the simulations we performed, and start
a new ensemble of simulations at the original density, but with
the more ``mature'' $\widehat{CDF}$.
The map from current void $CDF$ to future void $CDF$ is the
``coarse timestepper'' of the unavailable equation for the
macroscopic observable evolution; 
the composition of this map with the rescaling step constitutes
the ``renormalized coarse timestepper'', the main tool
of our dynamic renormalization procedure, schematically 
summarized in Fig. \ref{fig2}.
Given this map, several approaches to the computation
of the long-term coarse self-similar dynamics exist.
The simplest is successive substitution -- we repeat coarse
renormalized time-stepping again and again, and observe the
approach of the void $CDF$ to its self-similar shape {\it at
the scale we chose} (parametrized  by $\delta_0$).
The fixed point of this procedure lies on the group orbit of
scale invariance for the $CDF$ dynamics; it was selected through
a ``pinning'' or ``template'' condition (our choice of $\delta_0$).
This dynamic renormalization iteration, with $\tau = 4$ is shown in Fig. \ref{fig3}a, 
starting from a uniform void distribution with $\delta_0 = 0.2$; the inset shows the third
iteration before (dotted line) and after (solid line) rescaling.

Successive substitution will converge to {\it stable} self-similar 
solutions.
It is also, however, possible, to use {\it fixed point algorithms}
to converge to the self-similar void $CDF$ {\it at} $\delta_0$.
In effect, we wish to solve a functional equation for the fixed point
of the renormalized timestepper:
\begin{equation}
\widehat{CDF}(x,\tau) - CDF(x,0) = 0.
\label{fixp}
\end{equation}
Fixed point algorithms (like Newton's method) require the repeated
solution of systems of linear equations involving the Jacobian matrix of
some discretization of Eq. (\ref{fixp}).
Since we have no closed form equations for the void $CDF$ evolution,
this Jacobian is unavailable;
%not explicitly available; 
yet equation-free methods
of iterative linear algebra (like GMRES \cite{Kelley_Saad}) allow the
solution of the problem through a sequence of matrix-vector product
estimations.
In our case these estimations are obtained through the
``lift-run-restrict-rescale'' protocol performed at appropriately
selected nearby initial distributions (for details, see 
\cite{Manifesto}).
Short
bursts of stochastic simulation from nearby initial void distributions allows
us to estimate the action of the Jacobian 
%of the unavailable
%macroscopic Eq. (\ref{fixp}) 
on selected perturbations, and hence the
solution of linear equations and ultimately, of the nonlinear fixed point problem.
Fig. \ref{fig3}b shows the iterates of such a matrix-free fixed point
computation for Eq. (\ref{fixp}) through a Newton-Krylov GMRES algorithm applied to a 
100-point uniform finite difference discretization of the void $CDF$ in $x$;
our initial guess was a uniform void $CDF$ ($\delta_0 = 0.2$), and we converge 
to the self-similar shape within 
%6 
a few iterations; the inset shows the evolution of the norm
of the residual vector with iteration number.
{\it Lifting} from this coarse description (100 numbers) to realizations
of the $CDF$ (200,000 particles) involved linear interpolation of the $CDF$.
Relatively short runs ($\tau =1$) were used to construct the coarse
renormalized timestepper in Fig. \ref{fig3}b; its fixed
point is independent of the reporting horizon $\tau$.
%; for problems with power-law behavior in time, the
%critical exponents can be evaluated 
%upon convergence to the self-similar shape.

%
%It shoudl therefore be possible to construct 
%bifuration detection algorithms (wrapped around the coarse
%renormalized timesepper) that -given a good initial guess-
%will converge to the parameter boundary between
%``normal'' and self-similar dynamics.

\begin{figure}[htbp]
\centering
\leavevmode
\includegraphics[height = 2.6in]{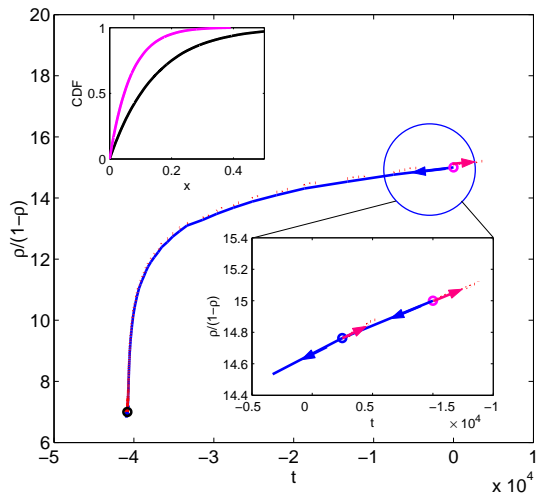}
\caption{Reverse coarse projective integration: the dotted pink line 
is the forward estimation of the rate; the solid blue line is the projection backward in time.
The upper inset shows (in the same color code as the large dots in the figure)
the CDF profile at the beginning and the end of the overall run.}
\label{fig4}
\end{figure}

Wrapping traditional 
numerical algorithms around coarse timesteppers
enables several computational tasks, of which the fixed
point computation above is only one example.
Another interesting example is the so-called coarse projective
integration; here, short runs of the stochastic
simulation are used to estimate the local time derivatives
of the macroscopic observables (e.g. of
the void $CDF$); these estimates are used to ``project'' the
void $CDF$ ``far" into the future.
%
%Saving of computational time 
%results for problems characterized with separation
%of time scales between fast and slow dynamics.
%
The simplest such algorithm is projective forward Euler;
see 
\cite{Manifesto} for multi-step and implicit coarse projective integration. 
%
%Once again, the important feature is that the
%microscopic simulations used to estimate the coarse time
%derivatives are {\it discarded}, and new simulations are
%started at the {\it projected} macroscopic state in the future.
%
One does not evolve the microscopic problem; one evolves the
coarse-grained closure of the microscopic problem, which is
postulated to exist, but is not explicitly available.
Here we apply coarse projective integration {\it backward in time} 
(as discussed in \cite{GearReverse} this
is appropriate for systems with large time scale separation
between fast stable dynamics and slow dynamics).
%
%In our case, we can use reverse coarse projective integration
%to enable our {\it forward in time} kMC to study the void $CDF$ 
%evolution {\it backward} in time.
%
Starting with the self-similar $CDF$ at some
(relatively large) density (see Fig. \ref{fig4} and its insets) we
lift to microscopic realizations, evolve briefly {\it forward} in time, 
and estimate the {\it rate} of void $CDF$ evolution.

This rate (estimated from the short pink run)
is used to project the void $CDF$ for a (longer) interval
{\it backward} in time (in blue).
The forward simulations are then discarded, and
new simulations are initialized at the ``earlier'' void $CDF$.
Repeating the procedure clearly shows an acceleration in backward 
time as the density becomes lower; the probability of large voids
becomes increasingly larger, the backward dynamics
evolves faster and faster and their fluctuations intensify,
making rate estimation from the short forward runs difficult.
Fig. \ref{fig4} seems to suggest a 
``reverse finite time" rate singularity. 
This is only a visual suggestion, however; as the density becomes
lower, 
%our {\it lifting} procedure (i.e. the construction of distributions 
%with prescribed density) will give rise to voids larger than 1 with
%increasing frequency. At these conditions, 
we do not expect an evolution
equation for the void CDF to provide a good model.
More detailed physical modeling of the void-filling process will
be required for the study of this backward dynamics at lower densities
to become meaningful.

It is this ``visual suggestion" of a backward explosion of the density
evolution rate that suggested fitting the data as a function of $t+t_0$;
we find that, if we initialize with the self-similar $CDF$ (at whatever
density), evolution data can be fitted almost perfectly with a straight
line {\it for the appropriate $t_0$} in Fig. \ref{fig1}.
Based on Eq. (\ref{delta}), the value of the term $B$ from all
of our trajectories suggests that at time $t+t_0 = 1$ the self-similar
$CDF$ density is $\sim 0.60$; that is, $t_0 - 1$ 
is the time that it
takes for a simulation {\it initialized with the self-similar void $CDF$
at density 0.60}  to evolve to the initial density of our computational 
experiments (our $t=0$).
When the initial condition is {\it different} than the self-similar 
$CDF$ (e.g. if it is a uniform distribution, see the dash-dot line in
Fig. \ref{fig1})  finding a good $t_0$
{\it does not} collapse the entire transient on a straight line -- we see,
however, that
the solution asymptotically approaches the self-similar one (dashed line).

In summary, in this paper we implemented equation-free, coarse-grained dynamic
renormalization (simulation forward and backward in time, as well
as fixed point computations) to study the evolution of the void $CDF$
for a model glassy compaction problem; we found this evolution to
be governed by apparent asymptotic self-similarity over the range of densities
we could reliably compute.
The procedure is, in principle,
capable of converging to both stable {\it and unstable} self
similar solutions, find similarity exponents when they exist,
as well as to quantify the fixed point stability
(by using matrix-free iterative
eigensolvers like Arnoldi).
%
%Upon convergence of the fixed point iteration, matrix-free iterative
%eigensolvers (like Arnoldi) can approximate the {\it coarse}
%eigenvalues and eigenvectors of the dynamic renormalization
%flow, quantifying fixed point stability.
%
In a continuation  / bifurcation context the algorithms can
be used to track 
self-similar solutions in parameter space, detect their
losses of stability and bifurcations; 
%and follow 
%new-born branches (see, for example, \cite{...}); 
of particular interest is the {\it onset}, in parameter
space, of self-similar dynamics, which will appear in our
formulation as a fixed point bifurcation \cite{template}.
Finally, coarse projective
integration (forward or reverse) can be performed in physical space 
(void $CDF$ evolution in time) or 
%-more appropriately for self-similar 
%problems, 
in renormalized space (renormalized
void $CDF$ evolution in logarithmic time).
We believe that the bridging of continuum numerical techniques with microscopic
simulations we illustrated here may be helpful in the coarse grained study  of glassy dynamics
through atomistic/stochastic models, especially when the dynamics depend
on {\it parameters} of the microscopic rules.

{\it Acknowledgments} IGK acknowledges the support of a NSF-ITR grant.
PGK acknowledges the support of NSF-DMS-0204585, NSF-CAREER and the 
Eppley Foundation for Research.

%\bibliography{glass}

\end{document}